\documentclass[rsi,
 reprint,
 amsmath,amssymb,
 aip,
 floatfix,
]{revtex4-1}

\usepackage[T1]{fontenc}
\usepackage[utf8]{inputenc}
\usepackage{graphicx}
\usepackage{dcolumn}
\usepackage{bm}
\usepackage{siunitx}
\usepackage{physics}
\usepackage{hyperref}
\usepackage{booktabs}
\usepackage{gensymb}
\usepackage{makecell}
\usepackage{xcolor}
\usepackage[version=4]{mhchem}
\usepackage[french,english]{babel}

\begin{document}

\title{Comprehensive In Situ Characterization of Amorphous Solid Water Films with COSPINU2 set-up: Optical Properties, Density, Morphology, and Restructuring Dynamics}

\author{Delfina Toulouse}
\email[]{delfina.toulouse@sorbonne-universite.fr}

\author{Japhar Michoud}
\author{Sasha Cryan}
\altaffiliation[Current address: ]{Université Paris-Saclay, CNRS, Institut d’Astrophysique Spatiale, Orsay, France}
\author{Arya M. Nair}
\author{Pascal Jeseck}
\author{Patrick Marie-Jeanne}
\author{G\'eraldine F\'eraud}
\author{Mathieu Bertin}
\author{Jean-Hugues Fillion}
\author{Xavier Michaut}
\affiliation{MONARIS, Sorbonne Universit\'e, CNRS, UMR 8233, Paris}

\date{\today}

\begin{abstract}
Amorphous solid water (ASW) dominates interstellar ice mantles, but its porosity, density, optical response, and mechanical stability depend sensitively on how the ice is grown. Here, we present the characterization capabilities of COSPINU2, a cryogenic ultra-high-vacuum set-up designed to prepare and diagnose water-ice films under controlled background and doser-assisted deposition conditions. The instrument combines a calibrated gas-handling line, near- and mid-infrared Fourier-transform infrared spectroscopy, HeNe interferometry, broad-beam optical imaging, and quadrupole mass spectrometry. We demonstrate how these complementary diagnostics provide a coherent description of ASW films grown between \SI{10}{K} and \SI{150}{K}. For background-grown ices, a dual-angle interferometric method is used to retrieve refractive indices near 632 nm, while near-infrared interference fringes provide thickness calibration. The results show that lower deposition temperatures and higher growth rates produce films with lower refractive indices and stronger dangling-OH infrared signatures, consistent with the formation of more porous ASW. We also show that pressure-based density estimates are unreliable in the background geometry because cryopumping decouple chamber pressure from the local substrate flux. For doser-grown films, optical interference imaging allows reconstruction of dome-shaped thickness profiles and estimation of effective densities. Finally, sudden restructuring events are observed in thick (above \SI{2}{\micro m}), low-temperature porous ASW films, evidenced by abrupt changes in interferometric contrast, optical scattering, and dangling-OH features. These results establish COSPINU2 as a versatile platform for quantitative water-ice characterization and provide the physical baseline required for future studies of infrared irradiation and optical constants of interstellar ice analogues.
\end{abstract}

\maketitle

\section{Introduction}\label{introduction}

Water ice is the most abundant solid constituent of interstellar dust-grain mantles \cite{boogert_observations_2015, mcclure_ice_2023}. At low temperature, typically below about \SI{100}{K}, vapor-deposited water forms amorphous solid water (ASW), whose microscopic morphology can range from compact to highly porous depending on the deposition temperature, flux, and angle, as well as the background pressure of the experimental chamber \cite{mate_morphology_2012,baragiola_water_2003,dohnalek_deposition_2003}. This morphology is not a secondary experimental detail but controls the internal surface area of the ice and therefore influences adsorption energies, diffusion pathways, trapping, desorption kinetics, and the reactivity of molecules relevant to astrochemical environments \cite{minissale_thermal_2022,carmack_pore_2022}.

The increasing quality of astronomical ice observations, notably with modern infrared observatories such as JWST, has renewed the need for laboratory data that are both quantitative and traceable to well-defined ice-growth conditions. Several laboratory set-ups have been developed to reproduce selected aspects of interstellar and planetary ice environments \cite{mitchell_porosity_2017,munoz_caro_new_2010,herczku_ice_2021,allodi_complementary_2013,hudson_ir_2017}. However, each set-up involves specific compromises between cryogenic performance, optical access, deposition geometry, spectroscopic sensitivity, mass-spectrometric diagnostics, and irradiation capabilities. As a result, optical constants, band strengths, porosity estimates, and density values measured in one apparatus cannot always be transferred directly to another without considering the way ice grows in that specific instrument \cite{westley_density_1998,bossa_porosity_2014,escribano_interstellar_2025,satorre_density_2025}.

This point is particularly important for the COSPINU2 (COnversion de SPIn NUcléaire) experimental set-up at MONARIS, Sorbonne Université. The set-up is designed to study quantum-state-resolved processes involving \ce{H2} and its isotopologues on water-ice surfaces, including adsorption, desorption, mobility, and nuclear-spin conversion. These processes are highly sensitive to the surface morphology and internal porosity. COSPINU2 is equipped for  infrared diagnosis that allow it to probe H$_2$ and its isotopolgues \textit{in situ} on icy surfaces\cite{michoud_influence_2023, hixson_experimental_1992, buch_preferential_1993, buch_rotational_1993} and to retrieve optical properties of laboratory-grown water ice. A reliable use of the apparatus therefore requires a preliminary answer to a simple but central question: what kind of water ice is actually produced under each deposition condition, and are its physical properties reproduced over several measurements?

The present article addresses this question from an instrumentation perspective. Rather than treating the characterization of ASW as a separate preliminary step, we use it as a benchmark of the COSPINU2 diagnostic chain. The main objectives are (i) to describe the experimental configuration used to grow and monitor water-ice films; (ii) to establish a refractive-index and thickness calibration for background-grown ASW; (iii) to combine infrared spectroscopy and optical imaging to evaluate the morphology, porosity indicators, and topology of background- and doser-grown films; and (iv) to identify growth conditions that trigger sudden restructuring of low-temperature porous ASW. This approach provides both a validation of COSPINU2 as an ice-characterization instrument and a physical baseline for future experiments on optical constants, infrared processing, and \ce{H2}/ice interactions.

\section{Experimental set-up}\label{experimental}

\subsection{Overview of the COSPINU2 apparatus}\label{overview}

\begin{figure}[ht]
    \includegraphics[width=\columnwidth]{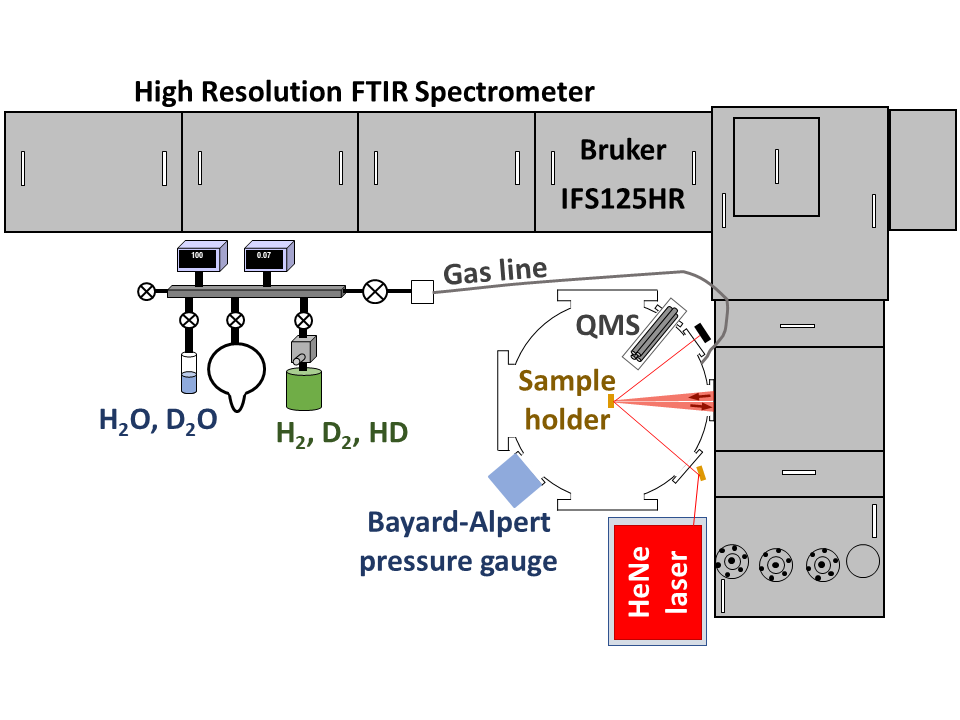}
    \caption{\label{Cospinu}Schematic layout of the COSPINU2 set-up showing the UHV chamber, closed-cycle cryostat, gas-handling line, FTIR spectrometer, HeNe laser and tunable mid-infrared laser paths, photodiode detection, optical-imaging access, and quadrupole mass spectrometer. The combination of these diagnostics enables simultaneous or sequential monitoring of ice growth, thickness, optical properties, morphology, and desorption.}
\end{figure}

\begin{figure}[ht]
    \includegraphics[width=\columnwidth]{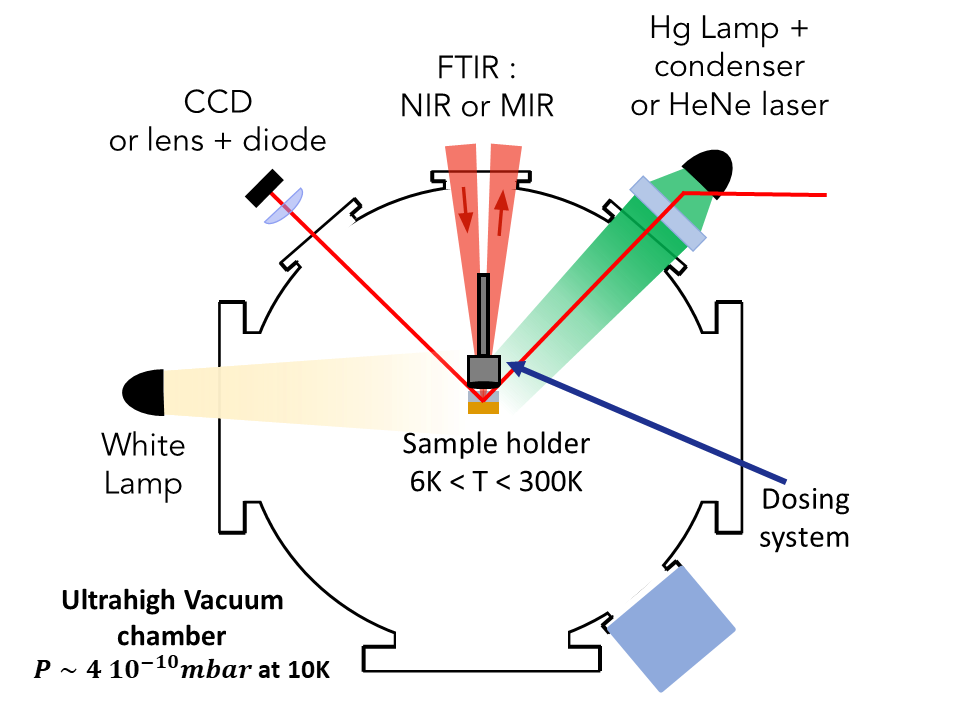}
    \caption{\label{Cospinu_enceinte} Detailed schematic of the vacuum chamber. The OFHC copper substrate is mounted on the cold finger of a closed-cycle cryostat. When the dosing system is retracted, the substrate can be viewed by the FTIR beam, the QMS (visible in figure \ref{Cospinu}), the HeNe laser monitored by a photodiode or alternatively the Hg lamp optical-imaging system recorded on a camera, and a white light at grazing incidence on the surface, also monitored with the camera.}
\end{figure}

COSPINU2 is an ultra-high-vacuum (UHV) cryogenic instrument designed for the controlled growth and \textit{in situ} characterization of interstellar ice analogues (figure \ref{Cospinu}). It couples a stainless-steel UHV chamber to a closed-cycle He cryostat holding an oxygen-free high-conductivity (OFHC) copper substrate (figure \ref{Cospinu_enceinte}). Optical access allows reflection-absorption Fourier-transform infrared (FTIR) spectroscopy at near-normal incidence, HeNe interferometry and mid-infrared laser irradiation at \SI{45}{\degree} incidence, and broad-beam optical imaging of the substrate using a green filtered monochromatic low pressure mercury lamp. A quadrupole mass spectrometer (QMS) monitors residual gases, deposition conditions, and desorption during temperature-programmed desorption (TPD) experiments.

\subsection{Main chamber and cryogenic substrate}\label{main}

\begin{figure}[ht]
    \centering
    \begin{minipage}{\columnwidth}
        \centering
        \includegraphics[width=\linewidth]{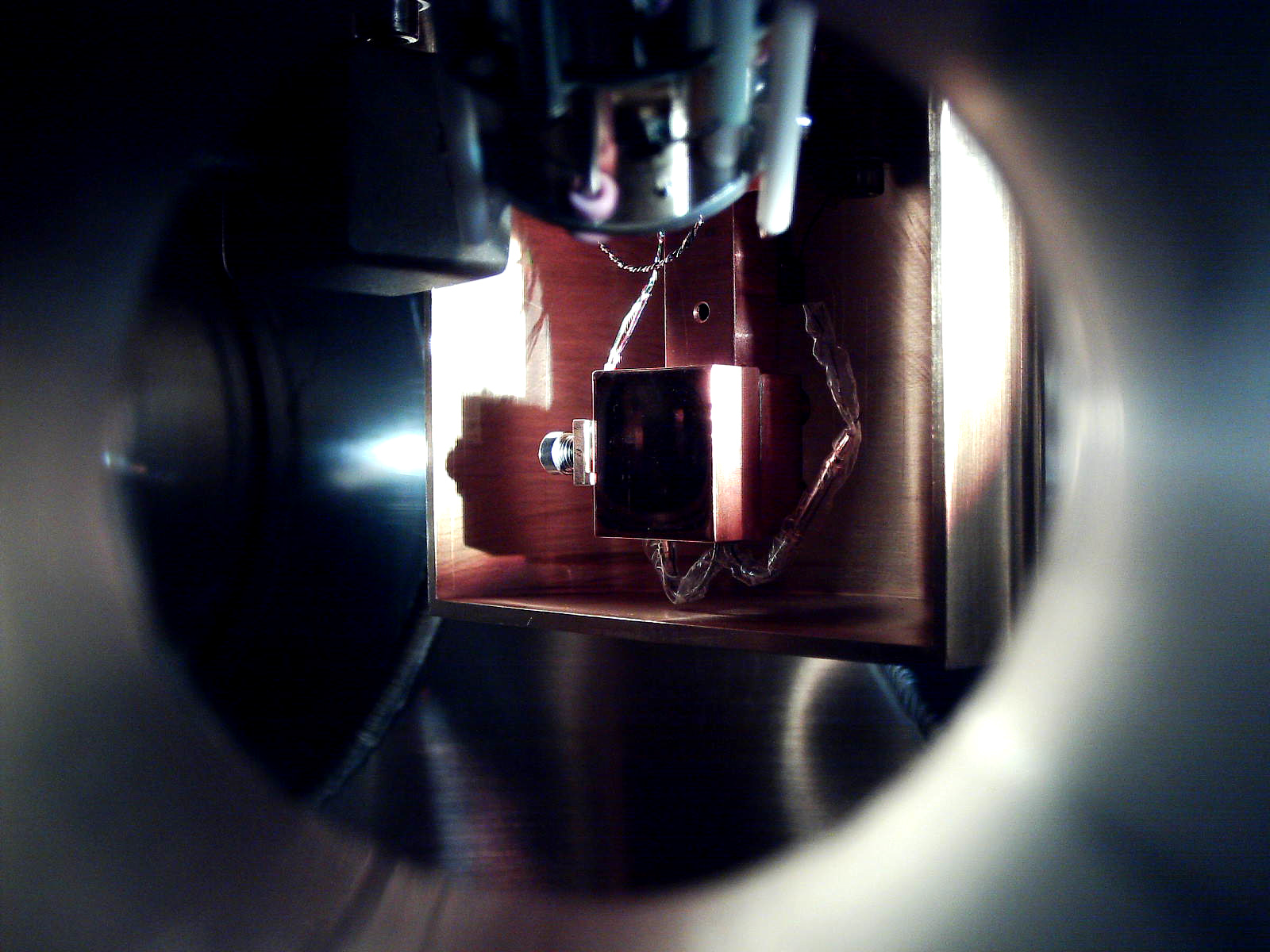}
    \end{minipage}
    \hfill
    \begin{minipage}{\columnwidth}
        \centering
        \includegraphics[width=\linewidth]{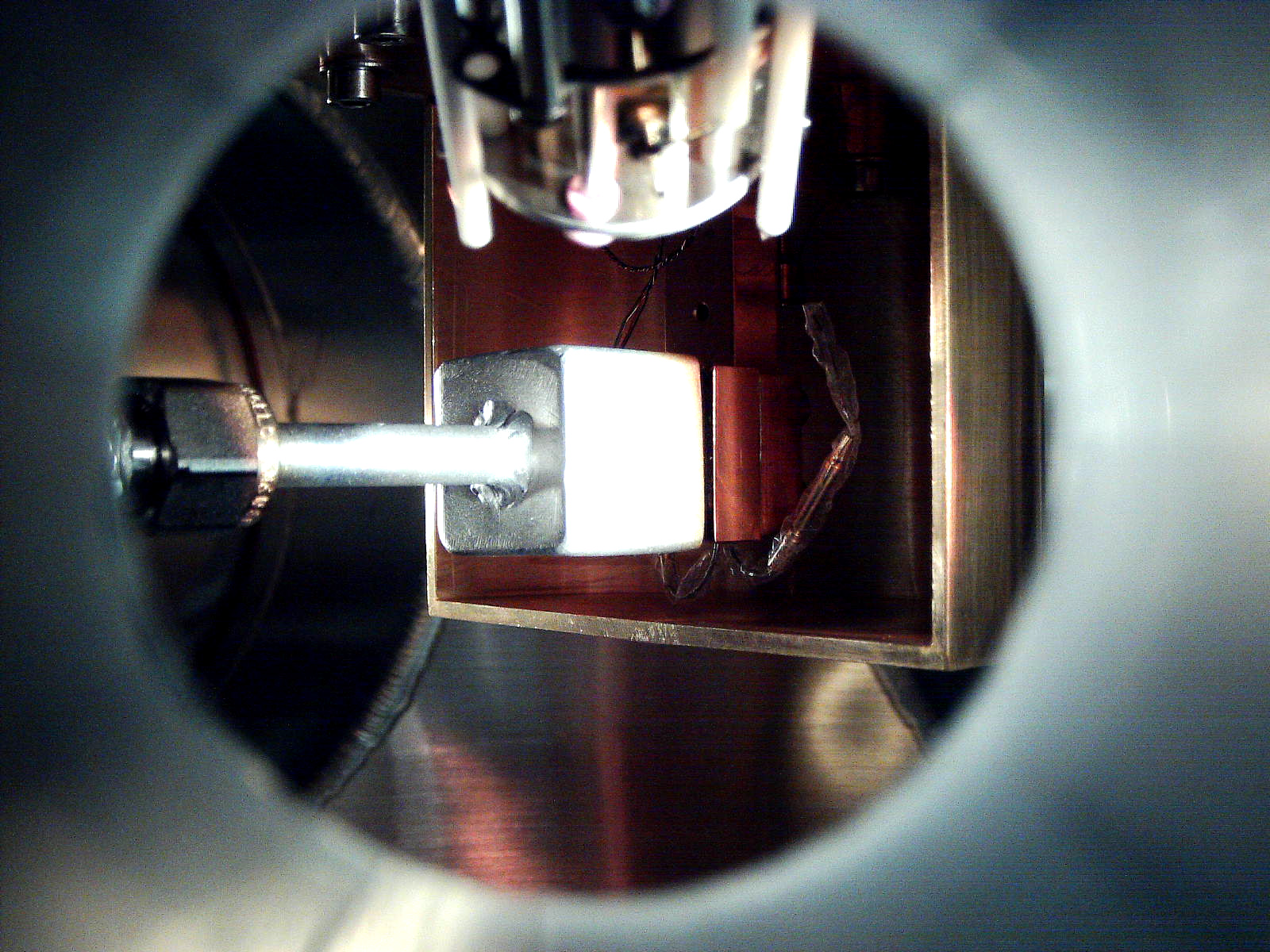}
    \end{minipage}
    \caption{\label{doseur}Photographs of the copper substrate without the doser (top) and with the doser placed in deposition position, approximately \SI{8}{mm} away from the surface (bottom). Background deposition leaves the substrate optically accessible during growth, whereas the doser enables localized deposition on the substrate but blocks direct optical monitoring during deposition.}
\end{figure}

The chamber is pumped by turbomolecular stages consisting of an Edwards nEXT400T turbo pump backed by an Edwards nEXT75 turbo pump, and reaches a base pressure between \SI{1e-10}{mbar} and \SI{4e-10}{mbar} at \SI{10}{K}. A Bayard-Alpert gauge (MKS 274042) measures the chamber pressure during operation. The water-ice films are deposited on a \SI{4}{cm^2} OFHC copper substrate mounted on a Cryomech PT405 closed-cycle cryostat, which provides \SI{0.5}{W} of cooling power at \SI{4.2}{K}. The minimum substrate temperature is \SI{5.8}{K}. This limit is mainly set by the geometry of the open cryoshield connected to the first stage of the cryostat and by the substrate design, shown in figure \ref{doseur}.

The substrate temperature is regulated by a Lakeshore 335 controller using a DT670 silicon diode sensor located close to the substrate. A second silicon diode sensor is connected to the cryostat head. Small oscillations associated with the cryocooler pumping cycle, at approximately \SI{1.4}{Hz}, induce temperature fluctuations of about \SI{1}{K} at the lowest temperatures. The substrate can be resistively heated using a cylindrical cartridge heater (Lakeshore, \SI{50}{\ohm}, \SI{50}{W}) located behind the surface. The thermal design limits heat transfer from the substrate to the cryostat head, ensuring that during linear heating ramps the substrate is warmer than the surrounding cryogenic parts. This separation is important for TPD measurements because it helps distinguish molecules desorbing from the substrate from molecules desorbing from other cold surfaces, such as cryostat tubes and the cryoshield.

\subsection{FTIR absorption spectroscopy}\label{FTIR}

\begin{figure}[ht]
    \includegraphics[width=0.95\columnwidth]{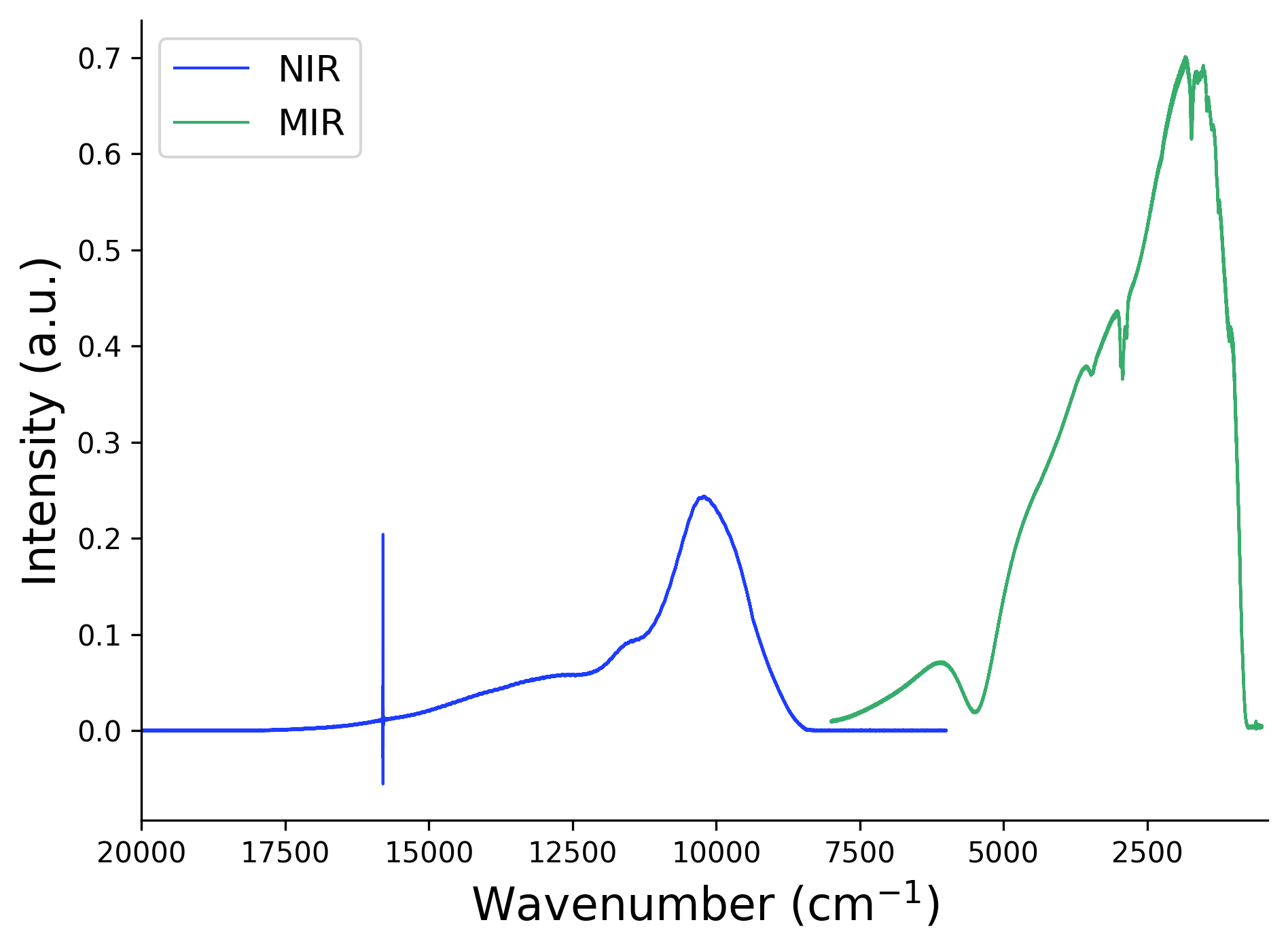}
    \caption{\label{Intensity_spectra} Spectra of the intensity of the beams in the two configurations used on the Fourier transform infrared spectrometer in this study. At \SI{15800}{cm^{-1}}, the NIR spectrum shows an intense line corresponding to the internal HeNe laser used for mirror alignment.}
\end{figure}

A Bruker IFS 125 HR spectrometer records reflection-absorption spectra of the ice-coated copper substrate at an incidence angle of the FTIR beam of \SI{5}{\degree}, going into the chamber through a BaF$_2$ window. This angle of incidence was evaluated by measuring the distance between the incoming and outgoing beams at a fixed distance from the substrate, and the error was estimated to be \SI{0.5}{\degree}. Two spectral ranges are used in this study, as visible in figure \ref{Intensity_spectra}. In the first range, addressed as near-infrared (NIR) hereafter as its maximum intensity is in the NIR, spectra are recorded from \SI{6000}{cm^{-1}} to \SI{20000}{cm^{-1}} using a tungsten source, a quartz beamsplitter (Bruker T402), and a silicon diode detector (Bruker SiD510). In the mid-infrared (MIR), spectra are recorded from \SI{500}{cm^{-1}} to \SI{8000}{cm^{-1}} using a globar source, a KBr beamsplitter (Bruker T301), and mercury-cadmium-telluride (MCT) detector (Bruker D313). The apertures of the diaphragm were set to \SI{1.5}{mm} for the NIR and \SI{0.5}{mm} for the MIR to get a maximum intensity on the spectra without saturating the detectors. This corresponds to spot sizes of about \SI{1.9}{mm} in the NIR and a little over \SI{0.6}{mm} for the MIR due to the magnifications from the optics inside the spectrometer. The NIR spot size was measured from pictures where the ice thickness is sufficient to diffuse the light and the spot becomes visible on the camera. The same magnification was applied to the MIR spot as it is not visible on camera. The errors are estimated to be \SI{0.1}{mm}.

NIR interference fringes provide thickness information as detailed in section \ref{refractive-index}, while MIR absorption bands probe the structure of the water-ice network as detailed in section \ref{infrared}. Although the spectrometer can reach a resolution of \SI{0.0025}{cm^{-1}}, all spectra were acquired at a resolution of \SI{0.5}{cm^{-1}} as a compromise between faster acquisition rate and good precision on the broad IR features of water ice spectra, with a boxcar apodization function and a zero-filling factor of 2. This acquisition mode provides a compromise between spectral resolution of the dangling-OH region and time resolution during ice growth with the background deposition technique described in part \ref{gas-handling}. To ensure reliable measurements of the mid-infrared absorbance spectra of strongly absorbing thick ice films, we first verified the linearity of the MCT detector up to an absorbance of 3 using germanium neutral density (optical density) filters (Edmund Optics infrared neutral density filters OD 0.3, OD 0.5, OD 1, OD 2 and OD 3).

\subsection{Quadrupole mass spectrometry}\label{quadrupole}

A Hiden DLS 10 QMS with an extended mass range up to 200~amu monitors the partial pressures of water vapor and residual gases during deposition. It is used to verify the cleanliness of the deposition environment, to follow the stability of the gas phase during ice growth, and to record temperature-programmed desorption of molecules of interest. During TPD experiments, the desorption flux of selected masses is monitored while the sample temperature is increased linearly with time until complete desorption of the ice.

\subsection{Optical diagnosis}\label{optical}

A \SI{632.8}{nm}, \SI{4}{mW} HeNe laser is directed onto the substrate at an incidence angle of \SI{45}{\degree}. The laser was aligned by centering two optical diaphragms on the entry and exit window of the beam, visible in figure \ref{Cospinu_enceinte}, and the error in precision is estimated to be a degree. Its intensity is measured on a photodiode, which allows for the characterization of ice thickness during and after deposition. The laser and its photodiode can be replaced by a monochromatic green extended light source and a camera, following the same path as the HeNe laser, as shown in figure \ref{Cospinu_enceinte}, to image the topology of the sample. The extended beam is made by filtering the green line at \SI{546.1}{nm} of an Hg vapor lamp with an interference filter. The resulting green beam is then extended with a condenser and centered on the BaF$_2$ entry window. Localized interference patterns on the ice allows the reconstruction of the macroscopic topography, as detailed in section \ref{characterization}. The HeNe laser and Hg lamp cannot be used at the same time and the switching from one to another implies realigning the entry path into the chamber each time. 

\subsection{Gas-handling line and deposition methods}\label{gas-handling}

Water vapor is introduced through a calibrated gas-handling line with a volume of \SI{1.8}{L}. The pressure inside the gas line is monitored through two capacitive Baratron pressure gauges (MKS-122AA) measuring up to 10 Torr for the first one and up to 1000 Torr for the second. The COSPINU2 set-up is capable of two different deposition geometries. First, during background deposition, water vapor is admitted into the whole chamber through a  6 mm stainless steel tube directed to the bottom of the chamber until a stable pressure set-point is reached either on the QMS or the pressure gauge. Assuming a sticking coefficient of unity, the collision rate on the substrate is estimated from gas kinetic theory.\cite{chiggiato_vacuum_2014} The number of collisions per cm$^2$ per second for a pressure P given in Pa is related to the temperature T and the mass m of the molecule, with k$_B$ the Boltzmann constant, via the formula :
\[
N_{col}= \frac{1}{4} P\times \sqrt{\frac{8}{k_BT\pi m}}
\]
For water, assuming that the temperature of the gas is homogeneous and equal to T=296 K, the deposition coefficient $n_{col}=N_{col}/p$ is :
\[
n_{col}(H_2O) = \SI{3.61e18}{collisions.Pa^{-1}.cm^{-2}.s^{-1}}
\]
Corresponding to :
\[
n_{col}(H_2O) = \SI{3.61e3}{ML.Pa^{-1}.s^{-1}}
\]
where one monolayer (ML) is defined as \SI{1e15}{molecules.cm^{-2}} (Note that for H$_2$, the calculation gives $n_{col}(H_2)$ = \SI{1.08e3}{ ML.Pa^{-1}.s^{-1}} with 1 ML($H_2$) = \SI{1e16}{molecules.cm^{-2}}). Integrating the chamber pressure during deposition therefore provides an operational estimate of the amount of water introduced under background-deposition conditions. In this geometry, ice grows on the substrate but also on other cold surfaces of the cryostat. The advantage of background deposition is that the substrate remains unobstructed at all times, allowing continuous FTIR and HeNe monitoring during growth. The growth rates of ices for the different pressures used for background deposition in this work are given in table \ref{vitesse_depot}.

\begin{table}
\caption{\label{vitesse_depot}Growth rates of water-ice films prepared by background deposition as a function of chamber pressure and substrate temperature. Rates are expressed in nm s$^{-1}$. A relative uncertainty of about 10\% is assigned to account for pressure fluctuations between experiments.}
\begin{ruledtabular}
\begin{tabular}{cccc}
 & \makecell{Fast\\\SI{4e-4}{Pa}}
 & \makecell{Medium\\\SI{6e-5}{Pa}}
 & \makecell{Slow\\\SI{6e-6}{Pa}} \\
\hline
\SI{10}{K}\rule{0pt}{2.6ex} & 2.5 & 0.4 & 0.04 \\
\SI{25}{K}  & 2.2 & 0.3 & 0.03 \\
\SI{90}{K}  & 2.0 & 0.3 &  \\
\SI{150}{K} & 1.5 & 0.2 &  \\
\end{tabular}
\end{ruledtabular}
\end{table}

\begin{figure}[ht]
    \includegraphics[width=\columnwidth]{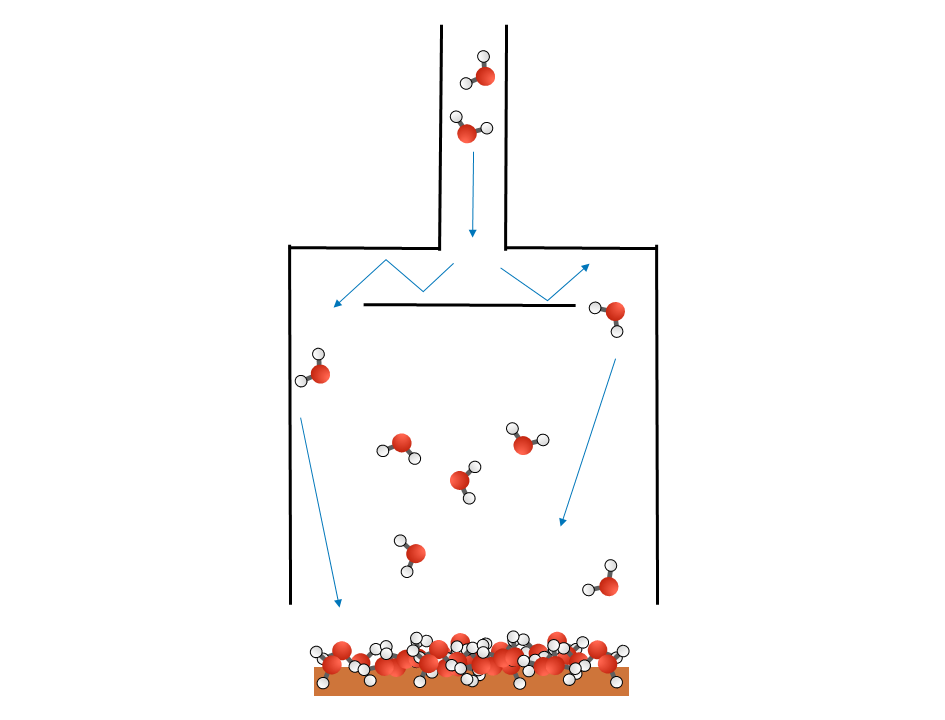}
    \caption{\label{dosing_system} Schematics of the interior of the dosing system. The molecules arriving from the gas line are deflected upon collision with the deflecting plate placed at the entrance of the doser.}
\end{figure}

In the second deposition geometry, a directed dosing system, hereafter referred to as a doser, that consists in a retractable tube ending with a large aperture adapted to the sample's size, through which the gaseous molecules can be injected using a UHV leak valve (Vacgen), is placed close to the substrate (figure \ref{doseur}). The tube is placed inside a metal bellows, and a stepper motor is used to move the doser in front of the substrate and to retract it after deposition without opening the chamber. The tube is made of stainless steel and specially designed for COSPINU2. Inside the large aperture, a deflective plate was placed right after the end of the metal tube so as to homogenize the trajectories of the incoming molecules from the cone of the directed gas flow (figure \ref{dosing_system}). This configuration produces a localized ice film and avoids the deposition of molecules on other cold regions within the UHV chamber. The number of water molecules introduced through the doser is determined from the pressure drop in the calibrated gas line. The main limitation is that the doser blocks the line of sight to the substrate during growth; therefore, FTIR spectra and optical images are acquired after deposition, once the doser has been retracted.

\begin{figure}[ht]
    \includegraphics[width=\columnwidth]{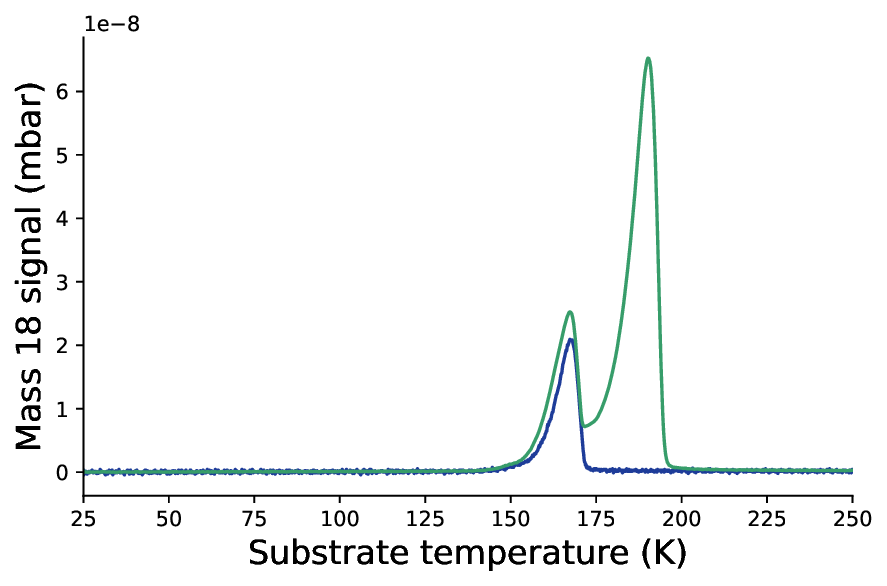}
    \caption{\label{TPD} TPD comparison of two water-ice films with similar thicknesses of approximately \SI{100}{nm}, deposited at \SI{25}{K} by background deposition (green) or with the doser (blue). The TPD experiments were recorded with a ramp of \SI{4}{K/min}. The low-temperature desorption peak observed in both traces is assigned to molecules deposited on the substrate. The additional, larger peak observed in the background deposition trace is attributed to water deposited on other cold cryogenic surfaces, demonstrating the substrate selectivity of the doser geometry.}
\end{figure}

 This configuration is beneficial because the QMS signal during TPD experiments can be directly linked to molecules desorbing exclusively from the copper substrate. This effect is visible on figure \ref{TPD}. Two TPD experiments were conducted after deposition of approximately \SI{100}{nm} of water ice at \SI{25}{K} and recorded with a ramp of \SI{4}{K/min} confirming the difference between the two deposition geometries. After background deposition, two desorption peaks are observed. The first one corresponds to water deposited on the copper substrate and desorbs at the expected temperature (from \SI{150}{K} to \SI{175}{K}). The second, larger peak is attributed to water molecules deposited on other cold parts of the cryostat. This confirms that background deposition does not selectively grow ice on the substrate alone, and that the temperature of the substrate is always higher than of the rest of the cold surfaces of the chamber. In contrast, doser-grown ice produces a single desorption peak associated with the substrate, showing that the doser strongly reduces deposition on other cold surfaces.

\section{Water-ice characterization}\label{water-ice}

\subsection{Refractive-index retrieval at 632 nm}\label{refractive-index}

\begin{figure}[ht]
    \includegraphics[width=\columnwidth]{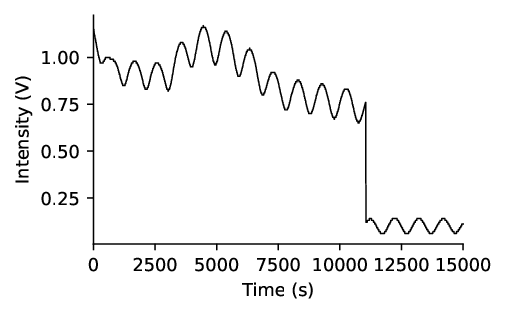}
    \caption{\label{interference_HeNe}HeNe reflected intensity recorded by the photodiode during background deposition at \SI{10}{K} and medium deposition rate ($P=\SI{6e-5}{Pa}$). The oscillations arise from interference in the growing ice film and provide the fringe period at \SI{45}{\degree} incidence. Slow baseline variations are attributed mainly to laser-power fluctuations. The abrupt loss of contrast near \SI{11000}{s} (thickness of approximately  \SI{3.8}{\micro\meter}) marks a restructuring event of the porous ice.}
\end{figure}

\begin{figure}[ht]
    \includegraphics[width=\columnwidth]{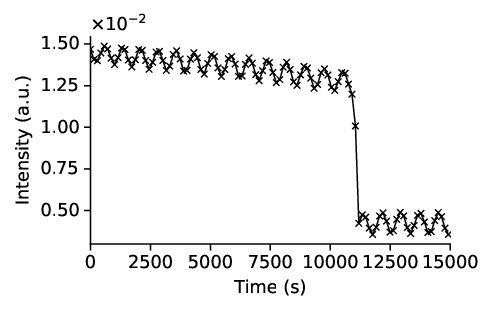}
    \caption{\label{interference_IR}Interference signal extracted from the NIR spectra around \SI{631.6}{nm} during the same background-deposition experiment at \SI{10}{K} and medium deposition rate ($P=\SI{6e-5}{Pa}$) as figure \ref{interference_HeNe}. Because the FTIR beam probes the ice at near-normal incidence, comparison with the HeNe period allows the refractive index to be retrieved using a dual-angle method. The discontinuity near \SI{11000}{s} (thickness of approximately \SI{3.8}{\micro\meter}) coincides with the restructuring event observed in the HeNe trace.}
\end{figure}

\begin{figure}[ht]
    \includegraphics[width=\columnwidth]{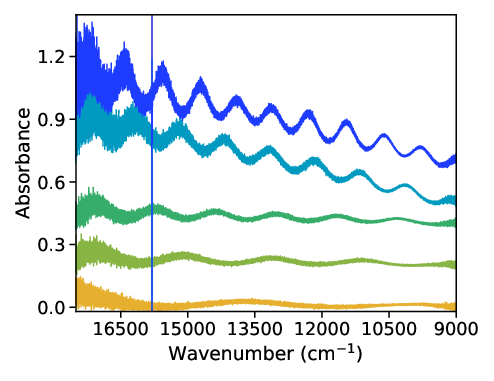}
    \caption{\label{spectres_NIR}Series of NIR spectra recorded during background deposition at \SI{10}{K} and medium deposition rate ($P=\SI{6e-5}{Pa}$), in increasing order of thickness from yellow to blue. The spectra are vertically offset for clarity. The periodic modulation with wavenumber is used to determine the film thickness after the refractive index has been obtained from the dual-angle analysis. The vertical line visible at \SI{15800}{cm^{-1}} corresponds to the saturation of the detector at the frequency of an HeNe laser placed inside the spectrometer and used for zero path difference calibration of the two mirrors.}
\end{figure}

During background deposition,  the reflected intensity of two probe beams were recorded as the ice film grows: on the photodiode monitoring the reflected HeNe beam at \SI{45}{\degree} incidence (figure \ref{interference_HeNe}) and in the FTIR NIR spectra by extracting the intensity of a line near \SI{632}{nm} from successive spectra, recorded during the ice growth, at \SI{5}{\degree} incidence (Figs.~\ref{interference_IR} and \ref{spectres_NIR}). The two measurements probe the same growing film but with different incidence angles. These two independent recordings display interference patterns. The period of these interference remains constant during the deposition (up to \SI{11000}{s} in figures \ref{interference_HeNe} and \ref{interference_IR}, with a period fluctuation of less than 1\%), indicating that the optical thickness of the ice increases linearly during time. Their combined analysis thus provides the refractive index without requiring prior knowledge of the film thickness.

At \SI{632.8}{nm}, the detector saturates because of a HeNe laser used for zero path difference calibration of the two mirrors inside the Michelson interferometer, which prevents a reliable direct analysis at the exact HeNe wavelength. We therefore averaged the spectral signal over \SI{20}{cm^{-1}} around \SI{631.6}{nm}. The wavelength difference relative to the external HeNe beam is only about 0.2\%, and is negligible compared with the uncertainty associated with fringe-period determination and beam angle.

During linear growth via the background deposition geometry (constant pressure in the chamber during deposition), the interference periods measured simultaneously with the HeNe laser at $\beta = 45\degree$ and with FTIR at $\alpha = 5\degree$ incidence provide two different optical paths. Denoting the period of the interference pattern by $\Delta t_1$ for the NIR spectra and $\Delta t_2$ for the HeNe laser, the ratio
\[
\gamma = \frac{\Delta t_1}{\Delta t_2}
\]
is related to the refractive index n at the wavelength of the two beams through:
\[
n^2 = \frac{\sin^2\beta - \gamma^{2}\sin^2\alpha}{1 - \gamma^{2}}
\]
This relation is equivalent to the dual-angle interferometric approach used for vapor-deposited water ice \cite{he_refractive_2022}, with a negligible correction from the small wavelength difference between the two signals.

The refractive indices obtained for background-grown ASW are reported in Table~\ref{indices_optiques}, where we observe two trends. First, at a given temperature, faster deposition generally produces a lower refractive index, consistent with a more porous film : molecules deposited more rapidly have less time to relax to more stable configurations or slightly rearrange before more molecules are deposited on top, leading to ices with more free space which in turn lowers the optical index, as was shown by Bossa et al.\cite{bossa_porosity_2015} Second, the refractive index tends to increase from \SI{10}{K} to \SI{90}{K}, consistent with enhanced molecular rearrangement and compaction at higher deposition temperatures. The values of the refractive index obtained at \SI{150}{K} must be interpreted cautiously because the ice structure is closer to crystalline water ice and because sublimation and growth kinetics become more sensitive to local thermal conditions. Overall, the COSPINU2 measurements agree with the recent view that refractive index and density depend strongly on deposition conditions. \cite{he_refractive_2022,escribano_interstellar_2025,satorre_density_2025}

\begin{table}
\caption{\label{indices_optiques}Refractive indices at approximately \SI{632}{nm} for background-grown water-ice films deposited at different chamber pressures and substrate temperatures. The values are retrieved from simultaneous HeNe and NIR interference periods. The larger uncertainty for fast deposition reflects the smaller number of spectra per fringe period.}
\begin{ruledtabular}
\begin{tabular}{cccc}
 & \makecell{Fast\\\SI{4e-4}{Pa}}
 & \makecell{Medium\\\SI{6e-5}{Pa}}
 & \makecell{Slow\\\SI{6e-6}{Pa}} \\
\hline
\SI{10}{K} & $1.17\pm0.10$ & $1.15\pm0.08$ & $1.27\pm0.08$ \\
\SI{25}{K} & $1.27\pm0.10$ & $1.25\pm0.08$ &  \\
\SI{90}{K} & $1.37\pm0.10$ & $1.37\pm0.08$ &  \\
\SI{150}{K} & $1.31\pm0.10$ & $1.39\pm0.08$ &  \\
\end{tabular}
\end{ruledtabular}
\end{table}

\subsection{Thickness calibration of background-grown ices}\label{thickness}

Once the refractive index is known, the full NIR spectra, exhibiting interference fringes as a function of the wavenumber, can be used to determine the absolute film thickness. At a fixed thickness, the spectral fringes visible in figure \ref{spectres_NIR} follow the interference condition:
\[
2nd \times cos(r)=m\lambda,
\]
where $d$ is the film thickness, $n$ is the refractive index, r is the refractive angle, $\lambda$ is the wavelength, and $m$ is the interference order. With an incidence angle of 5°, r is around 4° and the correction induced by the cos(r) factor is only 0.2 {$\%$}. We fixed the value of cos(r) to 1 for the following analysis. Assigning integer and half-integer orders to maxima and minima therefore allows the thickness to be reconstructed during and after deposition.

\begin{figure}[ht]
    \includegraphics[width=\columnwidth]{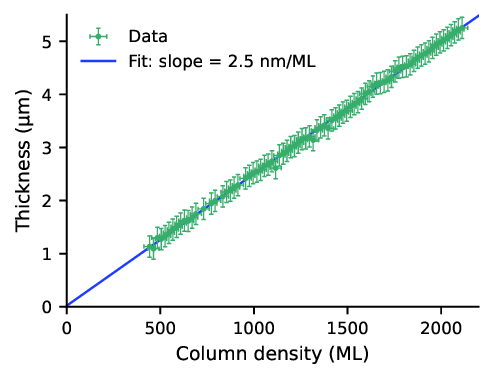}
    \caption{\label{density_bg}Thickness obtained from NIR interference fringes as a function of the column density estimated by integrating the chamber pressure during medium-rate ($P=\SI{6e-5}{Pa}$) background deposition at \SI{10}{K}. The linear fit provides an operational thickness calibration for COSPINU2 experiments performed in the MIR configuration, where NIR thickness fringes are not recorded simultaneously.}
\end{figure}

For typical background-deposition experiments, the film thickness reaches a few micrometers. The thickness retrieved from NIR fringes can be compared with the water column density inferred from integrated chamber pressure (figure \ref{density_bg}). This comparison provides an operational calibration between pressure exposure and ice thickness for experiments in which only the MIR configuration is used and NIR spectra are not available. Importantly, this calibration should not be interpreted as a direct density measurement, as discussed below.

\begin{figure}[ht]
    \centering
    \includegraphics[width=0.7\columnwidth]{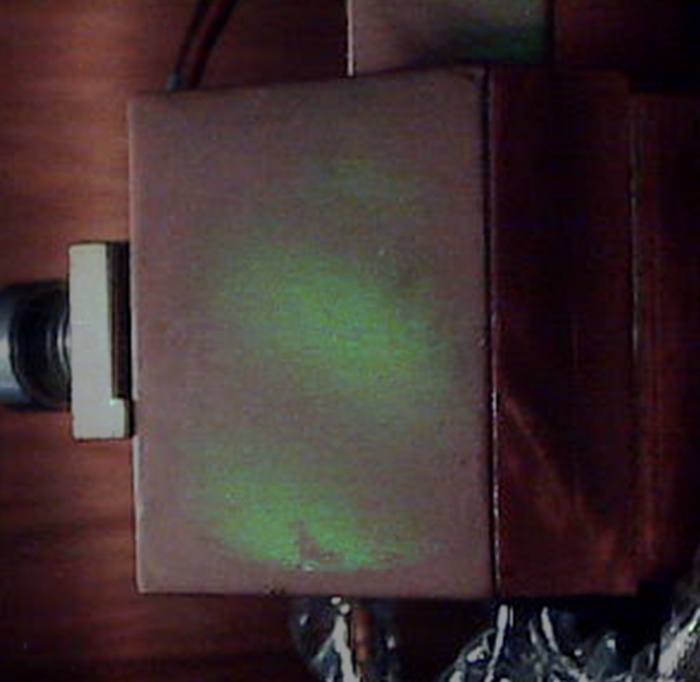}
    \caption{\label{gradient}Optical image of the substrate after background deposition of a water-ice film with a central thickness of approximately \SI{5}{\micro\meter}. For better clarity two pictures were superimposed : one of the substrate in white light, with a 40\% transparency effect, to clearly delimit the substrate, and one with the Hg lamp illumination. The interference stripes reveal a small macroscopic thickness gradient across the substrate. This gradient is much smaller than the characteristic size of the FTIR probing region and therefore does not dominate the local thickness determination.}
\end{figure}

Broad-beam Hg-lamp imaging of background-grown ice reveals weak interference stripes across the substrate (figure \ref{gradient}). These stripes indicate that the film is not perfectly homogeneous over the full \SI{4}{cm^2} copper surface, but instead displays a small diagonal thickness gradient. The corresponding slope is approximately $1.1\times10^{-3}$ degrees, or about 4 arcsec. This macroscopic gradient remains negligible over the FTIR spot sizes. The local optical measurements can therefore be treated as representative of the region probed by the spectrometer.

\subsection{Infrared signatures and limitations of pressure-based density estimates}\label{infrared}

The thickness calibration described above also reveals a limitation of pressure-based density estimates in COSPINU2. If the integrated chamber pressure is directly converted into the number of molecules deposited on the substrate, the inferred densities for background-grown ices range from approximately \SI{0.13}{g.cm^{-3}} for fast growth at \SI{10}{K} to \SI{0.20}{g.cm^{-3}} for medium-rate growth at \SI{150}{K}. These values are far below the expected densities of water ice, especially at high temperature where compact or crystalline structures should form. The discrepancy indicates that the chamber pressure does not provide a direct measure of the molecular flux reaching the substrate. Cryopumping by the cold cryostat parts and non-equilibrium gas distributions decouple the Bayard-Alpert pressure reading from the local growth rate. For background deposition, integrated pressure is therefore useful as a reproducibility parameter and as a thickness-calibration variable, but not as a standalone density diagnostic.
\begin{figure}[ht]
    \includegraphics[width=\columnwidth]{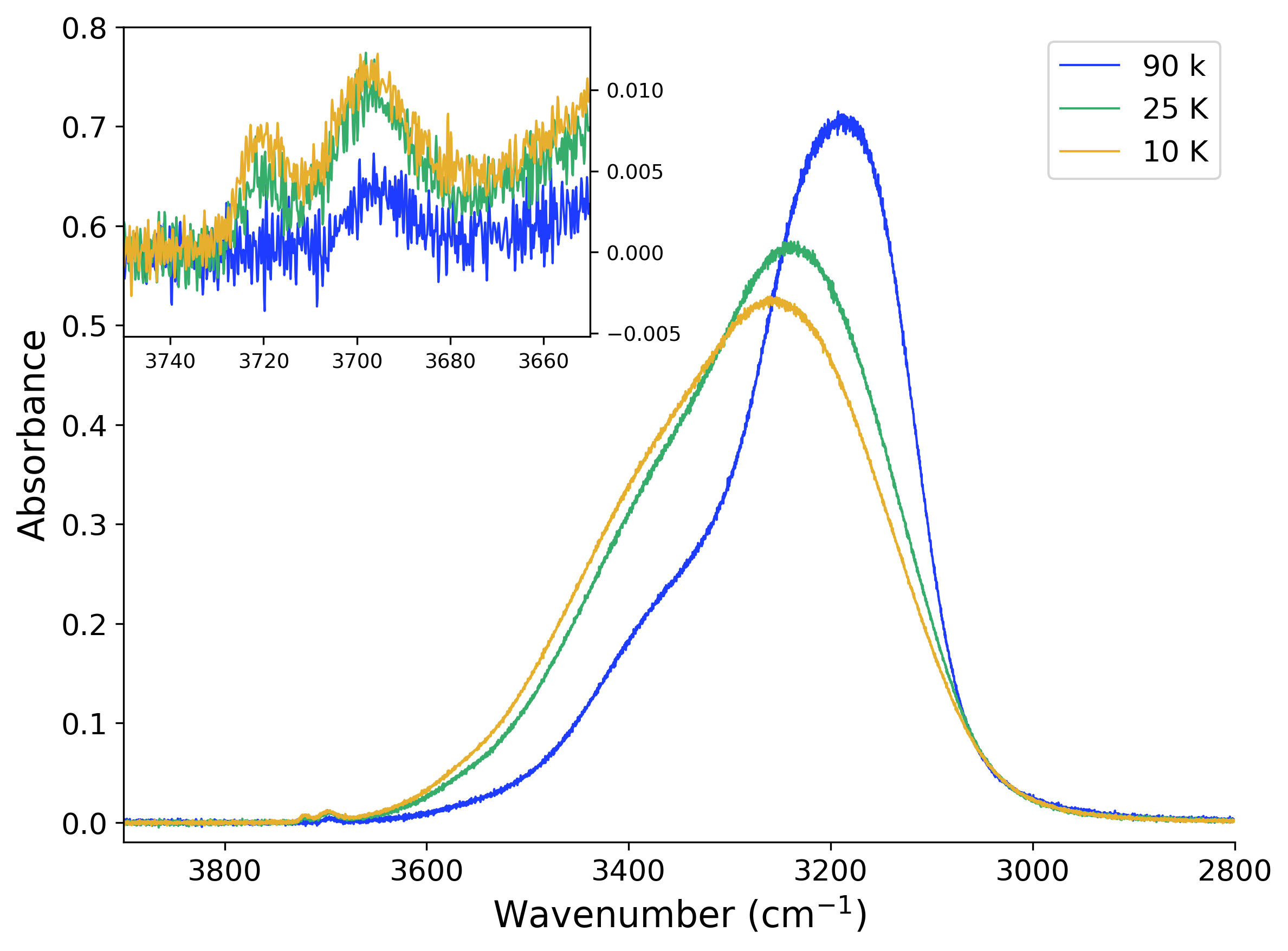}
    \caption{\label{temperature_effect}MIR spectra of the OH stretching band for ices of approximately \SI{350}{nm} deposited  at medium deposition rate ($P=\SI{6e-5}{Pa}$) using the background method at three different temperatures (\SI{10}{K}, \SI{25}{K} and \SI{90}{K}). The intense band around \SI{3300}{cm^{-1}} corresponds to bulk absorption, and small dangling-OH features corresponding to twice-coordinated water molecules (doubly -OH) and thrice-coordinated molecules (triply -OH) are visible at \SI{3720}{cm^{-1}} and \SI{3705}{cm^{-1}} respectively.}
\end{figure}

\begin{figure}[ht]
    \includegraphics[width=\columnwidth]{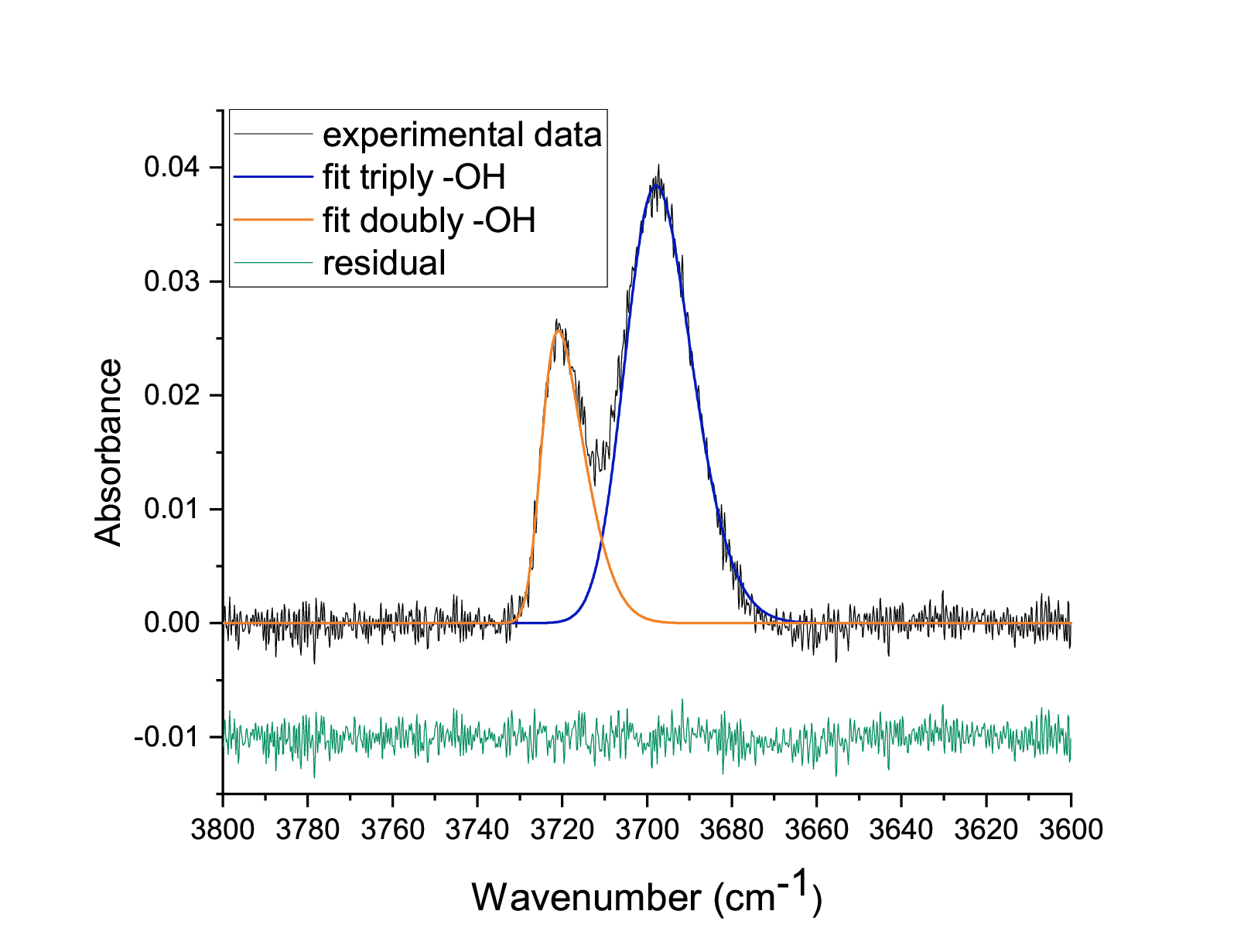}
    \caption{\label{fit_dangling}Example of the extraction of dangling-OH bands in the MIR spectrum. The spectrum presented here is obtained after polynomial correction for the baseline due to the presence in the vicinity of the strong absorption band of bulk water OH stretching. The under-coordinated OH features are fitted with a sum of 2 skew-normal distributions shown in blue and orange. The residual of the fitting procedure is shown below the spectrum in green and shifted by -0.01 for clarity. The integrated intensity of the doubly and triply coordinated dangling bands are measured through this fitting procedure during ice growth.}
\end{figure}

\begin{figure}[ht]
    \includegraphics[width=\columnwidth]{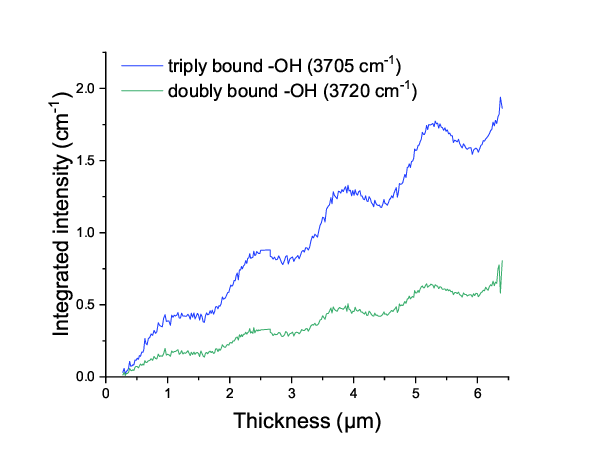}
    \caption{\label{doubly_triply_10K}Integrated areas of the doubly and triply coordinated dangling-OH bands as a function of ice thickness during slow background deposition ($P=\SI{6e-6}{Pa}$) at \SI{10}{K}. The global increase reflects the development of internal surface area in porous ASW, while the superimposed oscillations result from optical interference effects in the growing film.}
\end{figure}

MIR spectra provide a complementary probe of the microscopic structure of the water-ice network. Figure \ref{temperature_effect} compares the spectra obtained for water ices deposited at 3 different temperatures. The strong O-H stretch from H$_2$O molecules in the bulk is observed at \SI{3300}{cm^{-1}} and, at low deposition temperature, dangling-OH features are observed at \SI{3720}{cm^{-1}} (water molecules coordinated twice and called doubly -OH in the following) and \SI{3705}{cm^{-1}} (molecules coordinated thrice : triply -OH). These dangling bands assigned to under-coordinated OH groups at internal surfaces are commonly used as qualitative indicators of porosity \cite{bossa_porosity_2015,cholette_infrared_2009,raut_characterization_2007,buch_spectra_1991}. In COSPINU2, both dangling-OH bands are observed at \SI{10}{K} and \SI{25}{K}. At \SI{90}{K}, the triply coordinated feature remains detectable whereas the doubly coordinated band is no longer detected, consistent with partial compaction of the porous network. At \SI{150}{K}, both dangling features disappear within the present sensitivity, indicating a much more compact or crystalline structure.

\begin{figure}[ht]
    \includegraphics[width=\columnwidth]{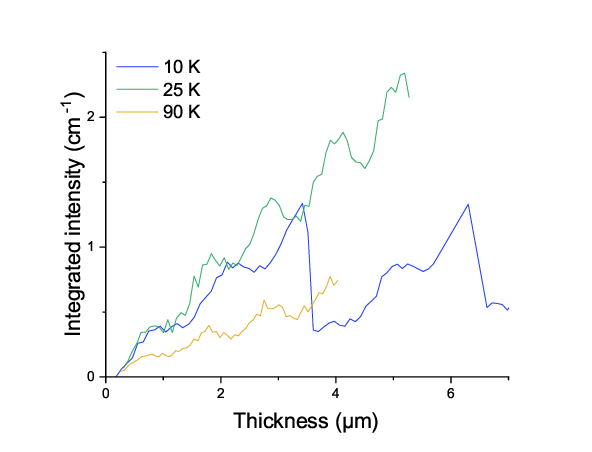}
    \caption{\label{triply_temperature}Integrated intensity of the triply coordinated dangling-OH band during medium-rate background deposition at \SI{10}{K}, \SI{25}{K}, and \SI{90}{K} ($P=\SI{6e-5}{Pa}$). The comparison highlights the progressive reduction of internal-surface signatures with increasing deposition temperature. The residual signal at \SI{90}{K} indicates that the ice is not fully equivalent to a compact, high-temperature film under the present growth conditions.}
\end{figure}

During growth at \SI{10}{K}, the integrated  intensities of the doubly- and triply- OH-dangling were determined by fitting the spectrum with the sum of two skew-normal distributions\cite{azzalini_skew-normal_2025} in order to take into account the asymmetry of the observed profiles, visible in figure \ref{fit_dangling}. The intensities of these two dangling bands generally increase with thickness (figure \ref{doubly_triply_10K}), consistent with the development of an extended internal surface in porous ASW. However, the intensities also show oscillations caused by interferences due to multiple reflections inside the ice, the same effects used previously for thickness determination. This prevents a direct conversion of dangling-OH integrated area into an absolute porosity. The MIR data should therefore be interpreted as a structural indicator that becomes quantitative only when combined with an optical model accounting for interference, thickness, and refractive index. Within this limitation, relative trends remain meaningful: dangling-OH signatures are stronger for lower temperatures  (figure \ref{triply_temperature}): ices deposited at \SI{10}{K} and \SI{25}{K} show a faster increase in intensity for the triply -OH band, indicating a faster growing internal area of the ice than at \SI{90}{K} and thus an increased porosity, and the doubly -OH band, present at \SI{10}{K} and \SI{25}{K} is no longer visible at \SI{90}{K}, suggesting a progressive reduction in internal surface of the ice as the deposition temperature increases, with a more compact structure. This effect is consistent with our measurement of the refractive index which suggested more compact ices at higher deposition temperatures. 

\subsection{Sudden restructuring of porous background-grown ASW}\label{sudden}

In several low-temperature background-deposition experiments, both the HeNe and NIR interference signals show abrupt discontinuities during growth (Figs.~\ref{interference_HeNe} and \ref{interference_IR}). These events appear as sudden losses of reflected intensity and fringe contrast. Because they are observed simultaneously with two independent optical paths, they cannot be attributed solely to a detector artifact. We assign them to rapid restructuring of a metastable porous ASW film, accompanied by a change in scattering and optical roughness.

\begin{figure}[ht]
    \centering
    \begin{minipage}{0.49\columnwidth}
        \centering
        \includegraphics[width=\linewidth]{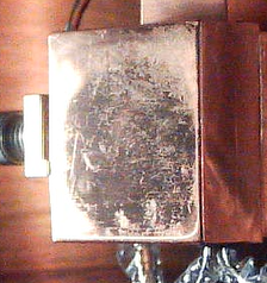}
    \end{minipage}
    \hfill
    \begin{minipage}{0.49\columnwidth}
        \centering
        \includegraphics[width=\linewidth]{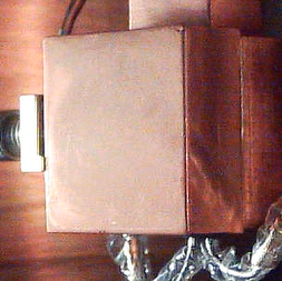}
    \end{minipage}
    \caption{\label{restructuration}White-light images of a \SI{10}{K} background-grown ice film before (left) and after (right) a restructuring event at medium deposition rate. The sudden whitening of the surface indicates enhanced light scattering, consistent with a rapid change in morphology, roughness, or internal cracking of the porous film.}
\end{figure}

Optical films recorded during deposition support this interpretation. A white light was added to illuminate the substrate as shown in figure \ref{Cospinu_enceinte} in such a way as to have a grazing incidence on the ice to see light scattering more clearly. At the time of restructuring, the surface becomes instantly whiter (figure \ref{restructuration}), indicating the sudden appearance of strongly scattering structures. This visual change coincides with the loss of interferometric contrast and therefore links the optical discontinuity to a real morphological modification of the ice.

\begin{figure}[ht]
    \includegraphics[width=\columnwidth]{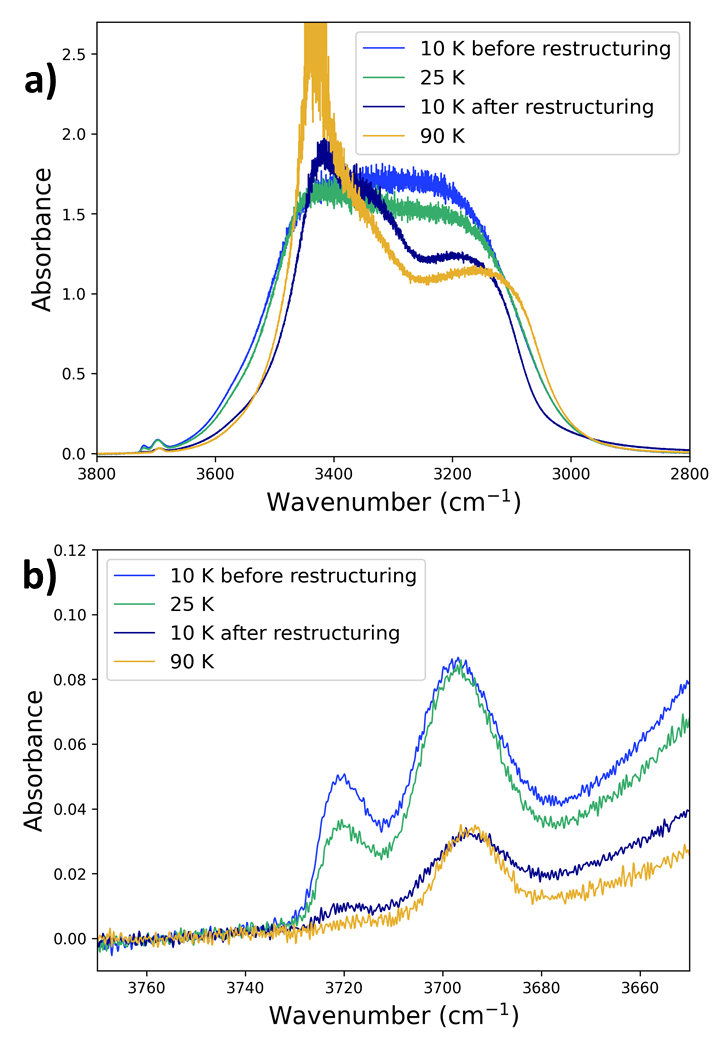}
    \caption{\label{dangling_restruc}Bulk stretching (figure a) and dangling-OH region (figure b) before and after restructuring of a medium-rate \SI{10}{K} background-grown ice, compared with spectra of \SI{25}{K} and \SI{90}{K} films of similar thickness grown at the same rate. The decrease and spectral evolution of the dangling-OH features after restructuring indicate a partial loss or reorganization of internal surface area.}
\end{figure}

The MIR spectra before and after restructuring further show that the event modifies the internal surface structure (figure \ref{dangling_restruc}). The bulk stretching band at high thicknesses combines the absorbance of the molecules and the interference effect of the probing beam inside the ice, distorting the shape of the band and making it quite sensitive to any changes within the ice. Before the event, the bulk stretching and the dangling-OH profiles resemble that of a highly porous low-temperature ice. After the event, the spectrum moves toward that of a more compact film, closer to the \SI{90}{K} reference. This does not necessarily imply complete crystallization; rather, it suggests a partial collapse, fragmentation, or reorganization of the porous network.

\begin{figure}[ht]
    \includegraphics[width=\columnwidth]{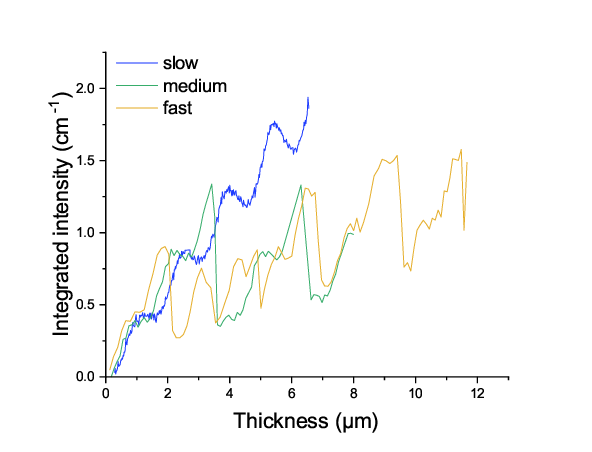}
    \caption{\label{10K_vitesse}Integrated intensity of the triply coordinated dangling-OH band during \SI{10}{K} background deposition at three growth rates. Slow deposition does not show abrupt restructuring within the explored thickness range. Medium-rate and fast deposition show recurrent discontinuities, indicating that the onset of restructuring depends on both local growth rate and accumulated ice thickness.}
\end{figure}
The dependence on deposition rate is shown in figure \ref{10K_vitesse}. At \SI{10}{K}, slow growth does not produce observable restructuring within the explored thickness range. Medium-rate deposition produces two events, around \SI{3.8}{\micro\meter} and \SI{6.2}{\micro\meter}. Fast deposition produces several events between about \SI{2}{\micro\meter} and \SI{12}{\micro\meter}. After each event, a new porous layer can grow on top of the restructured layer until another critical condition is reached. This recurrence suggests that the instability is not controlled by the copper substrate alone, but by the mechanical and morphological state of the growing ASW layer itself. Similar sudden changes have been discussed in relation to cracking, structural failure, and pore collapse in ASW \cite{bu_spontaneous_2016, may_release_2013, may_release_2013-1}, although the present observations occur during deposition at very low temperature rather than during warm-up or crystallization.

\subsection{Characterization of doser-grown ices}\label{characterization}

\begin{figure}[ht]
    \centering
    \begin{minipage}{0.49\columnwidth}
        \centering
        \includegraphics[width=\linewidth]{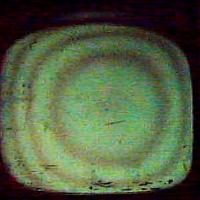}
    \end{minipage}
    \hfill
    \begin{minipage}{0.49\columnwidth}
        \centering
        \includegraphics[width=\linewidth]{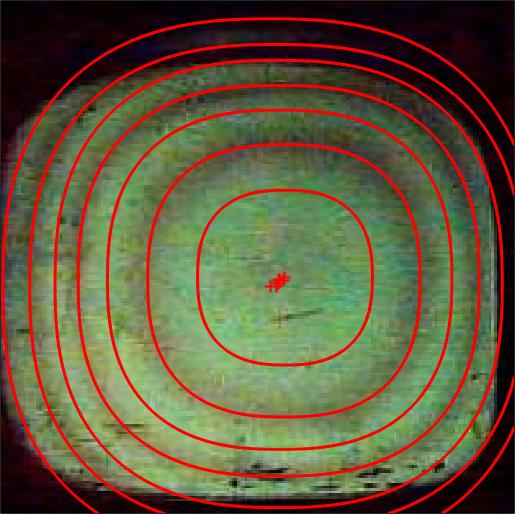}
    \end{minipage}
    \caption{\label{anneaux}Optical image of a doser-grown ice film after deposition. The raw image (left) shows Newton-ring-like interference patterns produced by the non-uniform thickness profile. The fitted image (right) shows the squircle-based isolines used to reconstruct the macroscopic thickness map of the deposited dome.}
\end{figure}

Doser-grown ices are characterized after deposition by combining NIR spectroscopy with broad-beam Hg-lamp imaging. In this configuration, the ice is localized near the center of the substrate and displays Newton-ring-like interference patterns (figure \ref{anneaux}). These rings reveal a dome-shaped thickness profile, with micrometer-scale thickness variations across the \SI{2}{cm} substrate. The dome results from the geometric distribution of molecules emerging from the doser, including the finite size of the doser opening and the deflective plate.

\begin{figure}[ht]
    \includegraphics[width=\columnwidth]{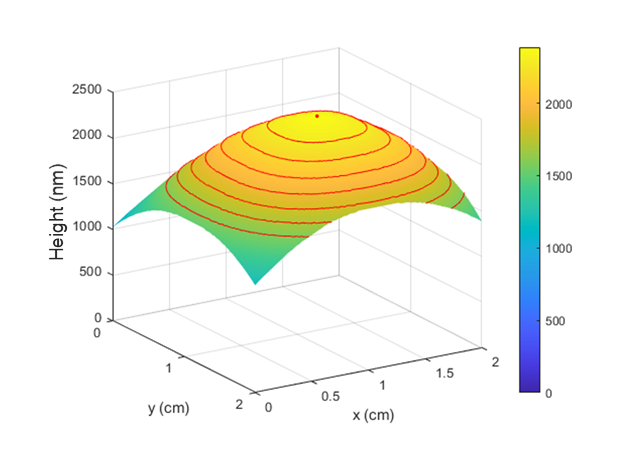}
    \caption{\label{volume}Example of a three-dimensional thickness map reconstructed for a doser-grown ice film. The map is obtained from optical interference isolines from the interference pictures (red lines) and anchored by the absolute NIR thickness measurement. This reconstruction provides the ice volume needed to estimate the effective density of doser-grown films.}
\end{figure}

The thickness reconstruction combines two pieces of information: the relative height isolines extracted from the optical interference image and the absolute thickness measured by NIR spectroscopy near the center of the substrate. Although the NIR spot is slightly offset from the maximum of the ice dome, it provides the absolute scale needed to convert the interference pattern into a three-dimensional thickness map (figure \ref{volume}).

\begin{figure}[ht]
    \centering
    \includegraphics[width=\columnwidth]{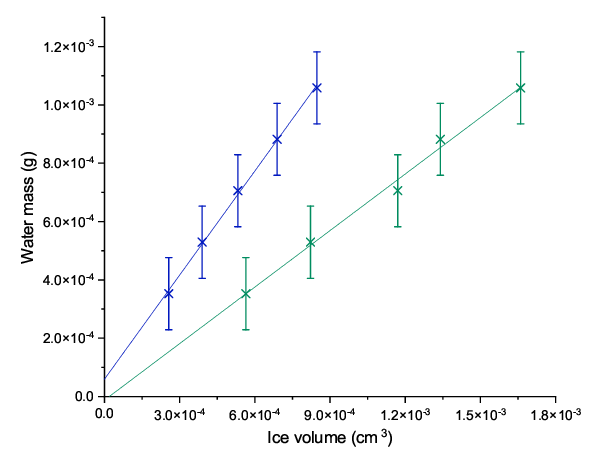}
    \vspace{0.5cm}
    \includegraphics[width=\columnwidth]{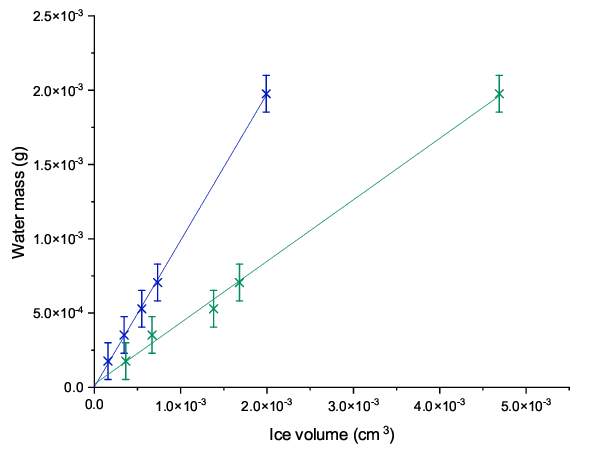}
    \caption{\label{densite_doseur}Water mass introduced through the doser as a function of reconstructed ice volume for two escape pressures from the doser during deposition: \SI{1e-7}{Pa} (top) and \SI{5e-7}{Pa} (bottom). Blue curves use only the volume contained within the substrate boundaries. Green curves include an extrapolated extended volume that accounts for molecules that would have landed outside the finite substrate area. The slope of the corrected curves gives the effective density of the doser-grown ice.}
\end{figure}

The amount of water introduced through the doser is obtained from the pressure decrease in the calibrated gas line using the ideal-gas law. Combining this molecular amount with the reconstructed ice volume gives an apparent density. If only the volume located on the substrate is used, the apparent density reaches about \SI{1.2}{g.cm^{-3}}, which is above the expected physical range for water ice under these conditions. This overestimate shows that not all molecules introduced through the doser are captured by the finite substrate. Some molecules are lost through lateral leakage or would have followed trajectories intersecting an area outside the copper surface.

To account for this geometric loss, the volume is extended beyond the substrate boundary by linearly extrapolating the outermost isolines until the thickness reaches zero. This corrected volume does not represent an actually measured ice volume on the substrate; instead, it estimates the volume that would have been formed if the collecting surface had been larger. With this correction, the density becomes \SI{0.63(4)}{g.cm^{-3}} for an escape pressure from the doser during deposition of \SI{1e-7}{Pa} and \SI{0.45(12)}{g.cm^{-3}} for \SI{5e-7}{Pa}, for films deposited at \SI{25}{K} (figure \ref{densite_doseur}). The lower density obtained at higher escape pressure is consistent with a more porous structure produced by faster growth as the density extracted from our experiments are consistent with those measured by Escribano et al.\cite{escribano_interstellar_2025}).

\begin{figure}[ht]
    \centering
    \includegraphics[width=\columnwidth]{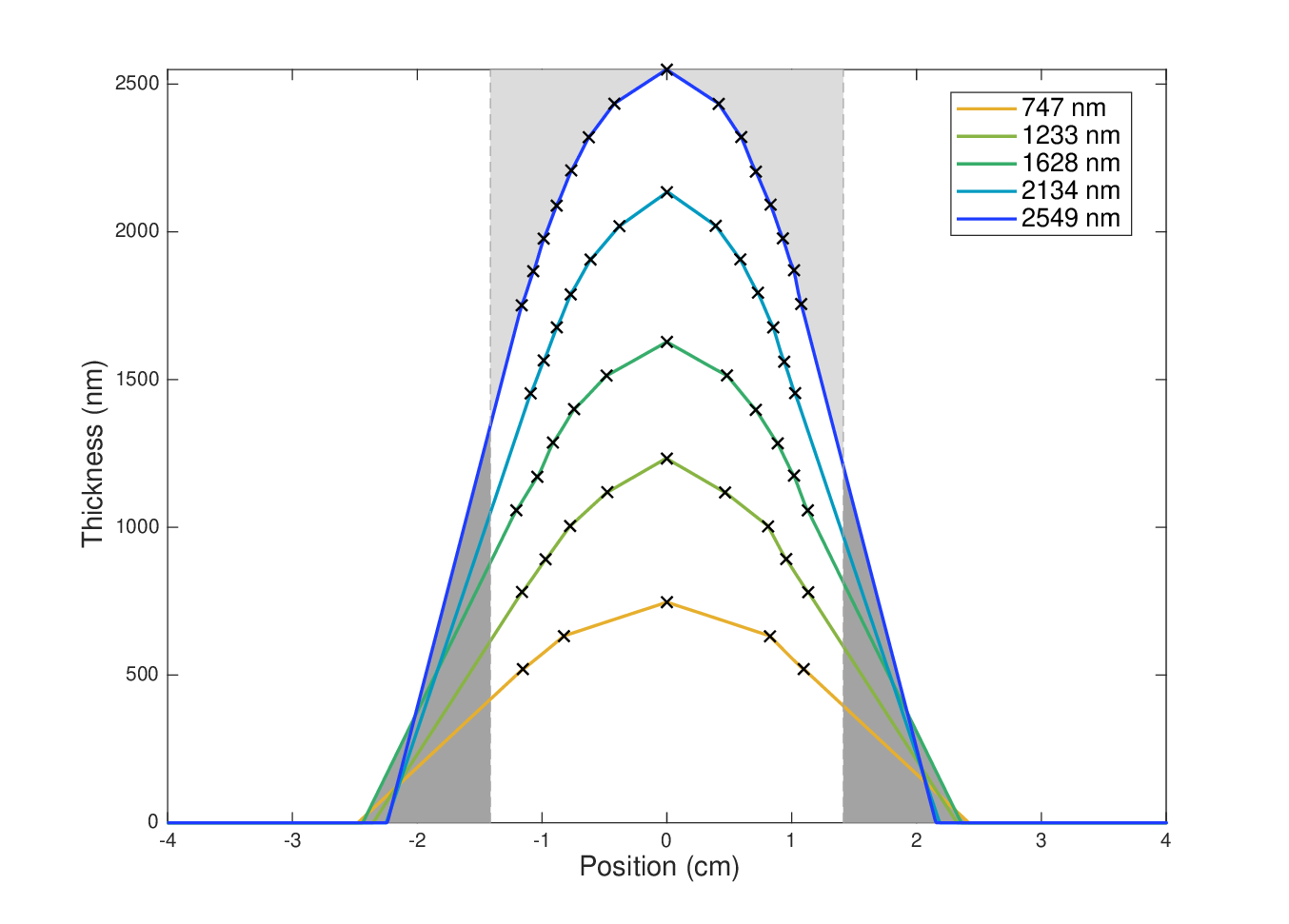}
    \caption{\label{profil}Thickness profiles along the substrate diagonal for successive \SI{25}{K} doser-grown films at an escape pressure of \SI{1e-7}{Pa}. Crosses correspond to experimental isolines extracted from the optical images. Dark gray regions indicate the extrapolated volume outside the substrate boundaries, while light gray regions show the profile expected for a background-grown ice of comparable maximum thickness.}
\end{figure}

Figure~\ref{profil} illustrates the evolution of the ices profile as the thickness increases. Successive doser depositions were made one by one on top of each other, always at the same escape pressure of \SI{1e-7}{Pa} of water at \SI{25}{K} to  progressively build an ice. Each time a NIR spectrum and an  interference picture were taken to reconstruct the 3D maps. The profiles were extracted from those maps across the diagonal of the substrate, and the extended volumes outside of the substrate boundaries used for density calculations are represented in dark grey zones. The figure also highlights the strong difference between an ideal background-deposited ice and our doser-deposited ices. Background deposition produces a nearly uniform film over the FTIR probing region, whereas doser deposition produces a localized dome with steep edges. This latter topology is advantageous when substrate-selective deposition is required, but it also means that thickness, density, and local molecular flux vary across the surface. Experiments that rely on doser-grown ice must therefore distinguish local thickness from total deposited volume. This thickness gradient is nevertheless negligible for MIR and NIR spectra when aligned on the center of the dome on the surface as the size of the spot is much smaller than the gradient (0.5 mm diameter for the MIR and 1.5 mm for the NIR).

\begin{figure}[ht]
    \centering
    \includegraphics[width=\columnwidth]{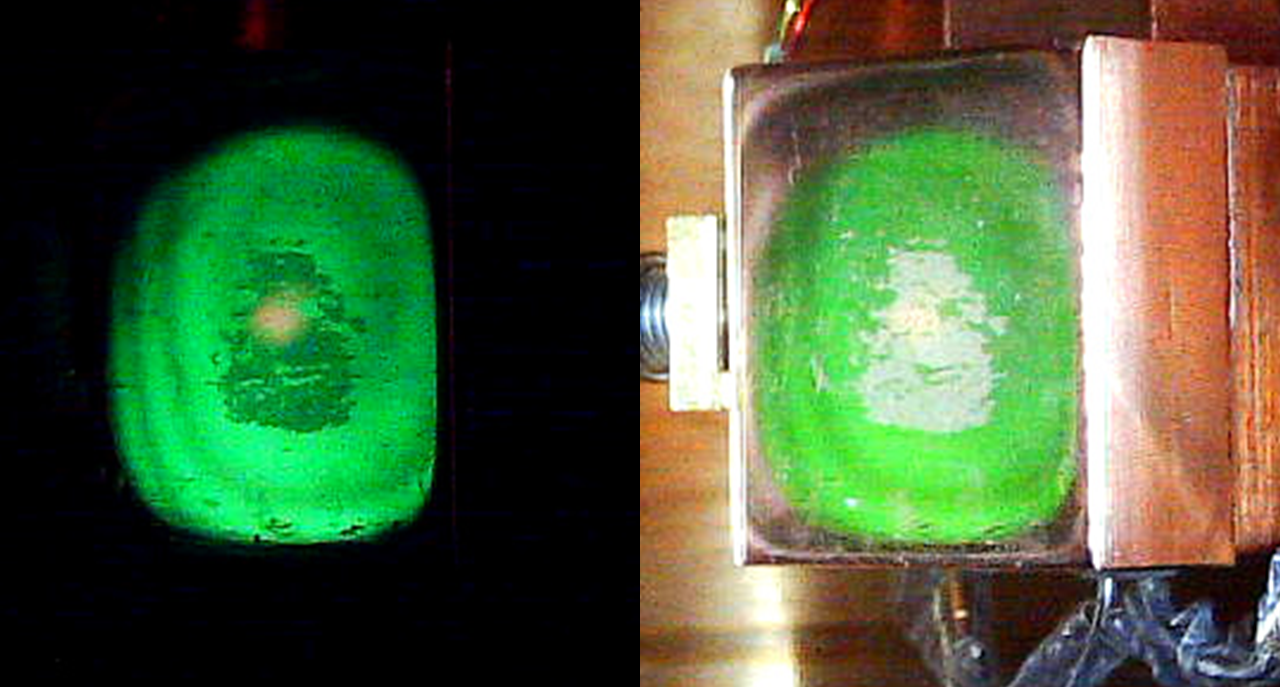}
    \caption{\label{tache_doseur}Optical image after doser deposition at low temperature. The NIR probing spot is visible near the center of the surface. The diffuse dark central region is attributed to localized restructuring of the thickest part of the doser-grown ice, analogous to the restructuring observed during background deposition but spatially confined by the inhomogeneous thickness and flux distribution.}
\end{figure}

At \SI{10}{K}, thick doser-grown films show localized dark regions under green illumination, mainly near the center where the ice is thickest (figure \ref{tache_doseur}). The same regions appear highly diffusive under grazing white-light illumination. We assign these features to the same restructuring phenomenon observed in background-grown ices. In the doser geometry, however, the effect remains localized because the thickness and incoming flux are strongly peaked near the center of the dome. The restructuring therefore appears to be controlled by local growth conditions rather than by a global temperature excursion or chamber-wide perturbation.

\section{Discussion}\label{discussion}

Overall, the present results establish a consistent link between deposition conditions and the optical, structural, and morphological properties of water-ice films grown in COSPINU2. The refractive indices obtained at 632 nm confirm that the optical properties of amorphous water ice are strongly dependent on the conditions under which the ice is grown. In the present experiments, the refractive index generally increases with deposition temperature and decreases when the deposition flux is increased (Table \ref{indices_optiques}). This behavior is consistent with the idea that low-temperature and fast-grown ASW films are more porous than higher-temperature deposits which allow a more efficient rearrangement of water molecules toward a denser structure. Similar trends have been reported in recent measurements of water ice analogues, where both the density and refractive index were shown to depend on deposition temperature and growth rate \cite{satorre_density_2025}. The present results are also consistent with previous studies showing that ASW morphology is highly sensitive to deposition geometry, molecular flux, and microscopic growth mechanisms \cite{dohnalek_deposition_2003,zhdanov_growth_2000}.

The low values of refractive index measured for the coldest deposits, especially at 10 K and 25 K, indicate that the films grown in COSPINU2 are highly porous under these conditions. This interpretation is supported by the MIR spectra, where dangling-OH features are clearly observed for low-temperature ices. These bands are commonly associated with  under-coordinated water molecules and internal surfaces, and are therefore useful qualitative indicators of porosity \cite{bossa_porosity_2015,cholette_infrared_2009,raut_characterization_2007}. However, the oscillations observed in the integrated dangling-OH intensities also show that these bands cannot be used as a direct quantitative measurement of porosity without accounting for optical interference effects in the case of thick films. The combination of interferometry and infrared spectroscopy is therefore essential to separate true structural effects from optical effects.

The comparison with the literature also highlights the importance of characterizing each experimental set-up independently. Westley et al\cite{westley_density_1998} reported density and refractive index values for vapor-deposited water ice films but observed only limited dependence on deposition temperature and rate. In contrast, more recent works show that porosity and refractive index can vary significantly with deposition conditions \cite{escribano_interstellar_2025,satorre_density_2025}. The differences between published datasets likely arise from variations in deposition geometry, background pressure, substrate configuration, and the method used to determine thickness and density. In COSPINU2, the column density calculated from the integrated background pressure is not fully reliable because cryopumping prevents the chamber from being in thermal equilibrium. Therefore, the refractive index and thickness calibration extracted from NIR spectra, and the MIR spectroscopic signatures provide a more robust description of the ice morphology than a column density directly inferred from pressure integration.

These results have direct implications for future optical-constant measurements in the MIR. Optical constants of amorphous and crystalline water ice are widely used to interpret laboratory and astronomical spectra \cite{hudgins_mid-_1993,mastrapa_optical_2008,mastrapa_optical_2009,kofman_refractive_2019,he_refractive_2022}. However, if the porosity and morphology of the laboratory ice are not well constrained, the derived constants may not be representative of the actual ice structure. This is particularly important for interstellar ice analogues, since band strengths, band positions, and optical constants can vary with porosity and temperature \cite{escribano_interstellar_2025,bossa_porosity_2014}. The COSPINU2 characterization therefore provides a necessary baseline for future retrieval of MIR optical constants using COSPINU2 under well-defined deposition conditions.

A second important result is the observation of sudden restructuring events during the growth of low-temperature ASW films. In background deposition experiments, this restructuring appears as an abrupt loss of the HeNe and NIR interference signals, together with a visible whitening of the surface under white-light illumination. The simultaneous change in optical response and surface appearance indicates that the effect is not only an optical artifact, but corresponds to a rapid modification of the ice morphology. The comparison of dangling-OH spectra before and after the event shows that the ice evolves from a highly porous structure, similar to that observed at 25 K, toward a more compact structure closer to the spectra obtained at 90 K. This suggests a partial collapse or reorganization of the porous network.

Similar sudden structural changes have been reported for ASW films in the context of cracking, pore collapse, and molecular release. Bu et al \cite{bu_spontaneous_2016}  showed that spontaneous cracking in ASW depends strongly on the microporous structure of the film, while May et al\cite{may_release_2013, may_release_2013-1} associated abrupt desorption events during heating with structural failure and crack formation in amorphous water ice overlayers. The present observations differ from these studies because the changes in absorption observed in the dangling bands cannot be solely explained by cracking, and the restructuring occurs during deposition at very low temperature, rather than during thermal crystallization or desorption.  The critical thicknesses before collapse reported by Bu et al\cite{bu_spontaneous_2016} are smaller than the ones we observed (they report L$_c$ = \SI{1}{\micro m} at \SI{10}{K} and normal incidence of the capillary array on the surface, and in our case we observe about \SI{2}{\micro m} for fast deposition and \SI{3.8}{\micro m} for medium deposition), however the comparison may not be so straightforward as it is hard to quantify the impact of the deposition method on the restructuring events. Their densities are also higher than what we estimated for our ices, highlighting again the difficulties of comparing two ices formed by two different set-ups. Nevertheless, the underlying mechanism may be related: a highly porous and mechanically fragile ice grows under conditions where incoming molecules continuously add stress and energy to a metastable structure. When the film becomes sufficiently thick or when the local incoming flux is high enough, the porous network may suddenly collapse, crack, and reorganize into a more compact and scattering structure.

The localized restructuring observed in doser-grown ices supports this interpretation. In these experiments, the ice thickness and deposition flux are not homogeneous over the substrate, and the restructuring is mainly observed near the center of the deposited dome, where the ice is thickest and the molecular flux is highest. The effect does not propagate over the entire substrate, which suggests that it is controlled by local growth conditions rather than by a global thermal event. This behavior is consistent with studies showing that ASW porosity evolves depending on temperature, thickness, and processing history \cite{bossa_thermal_2012,cazaux_pore_2015,carmack_pore_2022}. It also confirms that porous ASW is not a static material: even at low temperature, its morphology can evolve when mechanical stress, molecular flux, or external processing modifies the stability of the pore network.

Results of Bu et al\cite{bu_spontaneous_2016} indicate that the mechanical stability of ASW is governed by internal tensile stresses arising from the competition between surface forces acting on the internal pore walls and the adhesion of the film to the substrate. Using Griffith's fracture criterion\cite{griffith_vi_1921}, they estimated the tensile strength of ASW to be on the order of several tens of megapascals. Bu et al\cite{bu_effects_2016} also showed that these cracks profoundly modify the electrical properties of ASW. While thin homogeneous films lose most of their intrinsic surface potential during heating, cracked films retain significant electrical potentials because structural defects created by fracture act as efficient charge-trapping sites. These findings demonstrate that crystallization-induced cracking affects not only the mechanical integrity of amorphous ice but also its transport properties and electrostatic behavior. Collectively, these studies establish that the evolution of ASW is governed by a strong coupling between crystallization, microstructural reorganization, fracture mechanics, gas transport, and charge retention, with important implications for both laboratory experiments and astrophysical icy environments.

This restructuring has important practical consequences. It defines a limit for the reproducible preparation of highly porous ASW films in COSPINU2 and shows that high-flux, low-temperature deposition can produce unstable samples. The experimental limits for this effect were clearly identified, and all future experiments involving IR irradiation or H$_2$ adsorption and nuclear spin conversion will be conducted with water ices deposited at \SI{25}{K}, a temperature at which no restructuring was observed but the ice retained good porosity, and paying special attention to always align the FTIR and laser beams to the center of the domed ice produced by doser deposition to ensure minimum thickness gradient across the spot's diameter.

\section{Conclusion}\label{conclusion}

We have presented the characterization capabilities of the COSPINU2 set-up located at MONARIS, Sorbonne Université in Paris, France for amorphous water-ice films grown under cryogenic UHV conditions. By combining HeNe interferometry, NIR and MIR FTIR spectroscopy, broad-beam optical imaging, QMS, TPD, and calibrated gas handling, we determined refractive indices, thickness calibrations, morphology indicators, macroscopic topology, and effective density estimates for water-ice films grown between \SI{10}{K} and \SI{150}{K}.

For background-grown ices, the dual-angle interference method provides refractive indices near \SI{632}{nm}, while NIR spectral fringes yield absolute thicknesses. The measurements show that lower temperature and higher deposition rate favor lower refractive indices and stronger dangling-OH signatures, consistent with more porous ASW. At the same time, the comparison between optical thickness and integrated pressure shows that pressure-based density estimates are unreliable in the background geometry because cryopumping and deposition on other cold surfaces decouples the chamber pressure from the local substrate flux.

For doser-grown ices, optical interference imaging enables reconstruction of a dome-shaped three-dimensional thickness profile. After correcting for finite-substrate collection losses, the effective densities at \SI{25}{K} are \SI{0.63(4)}{g.cm^{-3}} at \SI{1e-7}{Pa} and \SI{0.45(12)}{g.cm^{-3}} at \SI{5e-7}{Pa}, again indicating that faster deposition produces more porous films. TPD confirms that the doser geometry strongly reduces ice growth on surrounding cold surfaces, making it suitable for substrate-selective experiments.

Finally, we observed sudden restructuring events in low-temperature porous ASW, both during background deposition and after doser deposition. These events are not caused by changes in temperature of the ice, but depend on deposition rate, local thickness, and flux, and are accompanied by changes in interferometric contrast, visible scattering, and dangling-OH signatures. They therefore define practical limits for reproducible preparation of thick porous ASW films, while also providing a sensitive probe of morphology-driven instability.

These results establish the physical baseline required for future COSPINU2 studies, including retrieval of mid-infrared optical constants, controlled infrared irradiation of ASW, and investigations of H$_2$ adsorption, desorption, mobility, and nuclear-spin conversion on water-ice surfaces.

\section*{Author Declarations}

\subsection*{Acknowledgements}
The authors gratefully acknowledge financial support for the development of the experimental set-up from the CNRS Terre \& Univers through the Commission Spécialisée Astronomie-Astrophysique ; the thematic action "Physique et Chimie du Milieu Interstellaire" (PCMI) of INSU programme national "Astro", with contributions from CNRS Physique and CNRS Chimie, CEA, and the CNES ; the Île-de-France DIM ORIGINES program ; the Initiative d'Excellence ISIM of Sorbonne Université ; Paris Observatory ; and the CNRS Chimie through PHOBEE Emergence project. Japhar Michoud acknowledges the Initiative d'Excellence ISIM of Sorbonne Université for funding his PhD fellowship.

\subsection*{Conflict of Interest}
The authors have no conflicts to disclose.

\subsection*{Author Contributions}

D. Toulouse wrote the manuscript. D. Toulouse, J. Michoud, S. Cryan and A. M. Nair carried out the experiments. D. Toulouse analysed the data. X. Michaut, P. Jeseck and P. Marie-Jeanne developed the experimental set-up. X. Michaut was in charge of overall direction and planning. M. Bertin, J.-H. Fillion and G. Féraud helped supervise the project. All authors provided critical feedback on the experimental results and the manuscript.

\subsection*{Data Availability Statement}

The data that support the findings of this study are available from the corresponding author upon reasonable request.

%
\makeatother

\end{document}